\documentclass{emulateapj}
\usepackage{url}

\begin{document}
\title{RADIO POLARIZATION OBSERVATIONS OF THE SNAIL: A CRUSHED PULSAR WIND
NEBULA IN G327.1$-$1.1 WITH A HIGHLY ORDERED MAGNETIC FIELD}

\shorttitle{RADIO OBSERVATIONS OF THE SNAIL PWN IN G327.1$-$1.1}
\shortauthors{Ma et al.}

\author{Y. K. Ma\altaffilmark{1}, C.-Y.\ Ng\altaffilmark{1}, N.
Bucciantini\altaffilmark{2,3}, P. O. Slane\altaffilmark{4}, B. M.
Gaensler\altaffilmark{5}, and T. Temim\altaffilmark{6,7}
} 
\altaffiltext{1}{Department of Physics, The University of Hong Kong, Pokfulam
Road, Hong Kong} 
\altaffiltext{2}{INAF --- Osservatorio Astrofisico di Arcetri, L.go E. Fermi
5, I-50125 Firenze, Italy} 
\altaffiltext{3}{INFN --- Sezione di Firenze, Via G. Sansone 1, I-50019 Sesto
F.no (Firenze), Italy} 
\altaffiltext{4}{Harvard-Smithsonian Center for Astrophysics, 60 Garden
Street, Cambridge, MA 02138, USA}
\altaffiltext{5}{Dunlap Institute for Astronomy and Astrophysics, The
University of Toronto, Toronto, ON M5S 3H4, Canada}
\altaffiltext{6}{Observational Cosmology Lab, Code 665, NASA Goddard Space
Flight Center, Greenbelt, MD 20771, USA} 
\altaffiltext{7}{CRESST, University of Maryland-College Park, College Park, MD
20742, USA}

\email{ncy@bohr.physics.hku.hk}

\begin{abstract}
Pulsar wind nebulae (PWNe) are suggested to be acceleration sites of cosmic
rays in the Galaxy. While the magnetic field plays an important role in the
acceleration process, previous observations of magnetic field configurations of
PWNe are rare, particularly for evolved systems.
We present a radio polarization study of the ``Snail'' PWN inside the supernova
remnant G327.1$-$1.1 using the Australia Telescope Compact Array.  This PWN is
believed to have been recently crushed by the supernova (SN) reverse shock.
The radio morphology is composed of a main circular body with a finger-like
protrusion.
We detected a strong linear polarization signal from the emission, which
reflects a highly ordered magnetic field in the PWN and is in contrast to the
turbulent environment with a tangled magnetic field generally expected from
hydrodynamical simulations. This could suggest that the characteristic
turbulence scale is larger than the radio beam size. We built a toy model to
explore this possibility, and found that a simulated PWN with a turbulence
scale of about one-eighth to one-sixth of the nebula radius and a pulsar wind
filling factor of 50--75\% provides the best match to observations. This
implies substantial mixing between the SN ejecta and pulsar wind material in
this system.  \end{abstract}


\section{INTRODUCTION} \label{sec:intro}
A massive star ends its life as a supernova (SN) explosion. This leaves behind
a supernova remnant (SNR) and sometimes creates a rapidly rotating pulsar. A
pulsar can produce an outflow of a magnetic field and relativistic particles
referred to as a pulsar wind. The interaction between such a pulsar wind and
the ambient medium, i.e.\ the SN ejecta for pulsars embedded in SNRs, results
in a termination shock, beyond which the shocked wind materials inflate a
broadband synchrotron-emitting bubble known as a pulsar wind nebula (PWN).

The evolution of PWNe can be divided into several stages
\citep[see][]{blondin01,vds01,vds04,gelfand09}. A PWN first expands
supersonically inside its associated SNR. The next stage starts when the SN
reverse shock, which is driven inward by the interaction between the ejecta and
the interstellar medium, crushes the PWN. The interplay between the shockwave
and the PWN is complex and will cause the PWN to reverberate
\citep[e.g.][]{vds01}. After the effect of the reverse shock fades away, the
PWN will expand subsonically into the SNR. Since a pulsar is generally born
with high space velocity, it can be significantly off-centered with respect to
the SNR during the reverberation stage and can protrude from its own PWN
because of the reverse shock interaction \citep{vds04}. This can result in a
complicated PWN morphology consisting of a ``relic'' component and an elongated
``head'' bridging between the relic and the pulsar \citep{vds04}. As the pulsar
continues to travel toward the edge of the SNR, its motion will eventually
become supersonic because the local sound speed in an SNR decreases moving
outward.  The ram pressure acting on the pulsar wind due to the pulsar's
supersonic motion will deform the PWN into a bow shock. For even older systems,
the pulsar can escape from its associated SNR shell and drive a bow-shock PWN
into the interstellar medium.

Broadband emission from PWNe can be observed from the radio to $\gamma$-rays.
In the radio regime, a PWN emits through the synchrotron process, and the
emission is often highly linear polarized because of the ordered magnetic field
configuration. A radio PWN is characterized by a centrally filled morphology
and a flat spectrum, with a spectral index of  $-0.3 \leq \alpha \leq 0$
($S_\nu \propto \nu^\alpha$). The synchrotron spectrum can extend to the X-ray
band where the spectrum is usually steeper than that in the radio due to
synchrotron cooling. GeV and TeV $\gamma$-ray emission from PWNe has also been
detected, which is due to inverse-Compton scattering \citep[see the review
by][]{gaensler06}.

One interesting area to explore in the study of PWNe is the interaction with
the SN reverse shocks. Theoretically, hydrodynamical (HD) modeling shows that
the interplay can give rise to turbulence in PWNe \citep{blondin01,vds04}.
However, the role of the magnetic field is unclear since magnetohydrodynamic
(MHD) efforts are lacking. Observationally, we aim to probe the magnetic field
structure of these PWNe to expand the currently inadequate sample size. The
results from our study can serve as inputs to MHD modeling.

One of the few examples of such systems is \object{G327.1$-$1.1}, which is an
SNR system containing a PWN that is believed to have been recently crushed by
the reverse shock. It was discovered as a non-thermal radio source
\citep{clark73,clark75}. Its peculiar radio morphology was revealed by the
Molonglo Observatory Synthesis Telescope (MOST) SNR survey \citep{whiteoak96}.
The SNR shell is $17^\prime$ in diameter and contains an off-centered PWN. The
latter consists of a circular main body with a $3^\prime$ diameter together
with a $2^\prime$-long finger-like structure protruding northwest from the main
body.

The first X-ray detection of G327.1$-$1.1 was reported by \cite{lamb81} using
the \textit{Einstein Observatory}. Further X-ray studies were conducted with
\textit{ROSAT} \citep{seward96,sun99}, \textit{ASCA} \citep{sun99},
\textit{BeppoSAX} \citep{bocchino03}, \textit{Chandra} \citep{temim09,temim15},
and \textit{XMM-Newton} \citep{temim09}. The X-ray remnant consists of a
diffuse thermal component covering the radio shell and a non-thermal component
coinciding with the radio PWN. An X-ray compact source at the tip of the finger
was discovered along with two prong-like structures extending northwest from
the finger into a diffuse, arc-like structure \citep{temim09,temim15}. The
compact source is interpreted as the neutron star powering the PWN, but the
nature of the prongs and the arc is not well understood. \cite{sun99} estimated
the distance to G327.1$-$1.1 to be $d = 9.0\,{\rm kpc}$ based on an empirical
relationship between the X-ray column density $N_{\rm H}$ and the color excess
$E(B-V)$, and the observed relationship between $E(B-V)$ and the distance $d$
along the line of sight to G327.1$-$1.1. We will adopt this number ($d =
9.0\,{\rm kpc}$) in our analysis.

In $\gamma$-rays, a preliminary report of TeV detection of G327.1$-$1.1 with
H.E.S.S. was described by \cite{acero12}. As the photon index in TeV was found
to be similar to that in keV, they suggested that the emission in the two bands
could originate from the same population of particles through inverse-Compton
and synchrotron processes, respectively. This source has not been detected in
the GeV range by \textit{Fermi} LAT \citep{acero13,acero15}.

By considering the current position of the presumed neutron star relative to
the SN shell, \cite{temim09} suggested that the neutron star is moving
northward. The main body of the PWN is interpreted as the relic, and the
overall PWN morphology could be caused by the reverse shock first crushing the
PWN from the northwest, thus pushing the relic to the southeast. The asymmetry
of the reverse shock interaction is attributed to both the inhomogeneous
ambient interstellar medium and the pulsar's motion. This picture is supported
by recent HD simulations \citep{temim15}. It is believed that the PWN has not
yet evolved into the bow-shock stage \citep{vds04,temim09,temim15}.

In this paper, we present radio observations of the PWN in G327.1$-$1.1 with
the Australia Telescope Compact Array (ATCA). We obtained polarimetric
measurements at 6 and 3\,cm to investigate the magnetic field structure of the
PWN. The observations and data reduction procedures are described in
Section~\ref{sec:obs}. We show the results in Section~\ref{sec:results} and
discuss our findings in Section~\ref{sec:discussion}. We close by summarizing
our work in Section~\ref{sec:conclusion}.

\section{OBSERVATIONS AND DATA REDUCTION} \label{sec:obs}
Radio observations of the PWN in G327.1$-$1.1 were performed with ATCA at 6 and
3\,cm on 2008 December 09 and 2009 February 19 with array configurations of
750B and EW352, respectively. The observation parameters are listed in
Table~\ref{table:para}. All of the observations were carried out prior to the
Compact Array Broadband Backend (CABB) upgrade \citep{cabb}. The data were
taken in continuum mode with all four Stokes parameters recorded over a usable
bandwidth of 104\,MHz at each frequency. We observed for a total integration
time of $20.8\,{\rm hr}$ in each band as a five-pointing mosaic. PKS
B1934$-$638 was adopted as the primary calibrator to set the flux density
scale, and PKS 1613$-$586 was observed at 20 minute intervals to determine the
antenna gain solution. We also processed archival ATCA pre-CABB data at 20 and
13\,cm, for which the observation parameters are also listed in
Table~\ref{table:para}. The 13\,cm observations were performed in continuum
mode with the same usable bandwidth of 104\,MHz. However, the 20\,cm
observations were performed in spectral line mode on the H~{\sc i} line, such
that only the total intensity (i.e.\ no polarization) within a narrow usable
bandwidth of 4\,MHz was recorded. The simultaneous observations at 20 and
13\,cm consist of a single pointing at G327.1$-$1.1, with array configurations
of 1.5D and 750D for a total integration time of $19.0\,{\rm hr}$ per band. PKS
B1934$-$638 was also used as the primary calibrator for these two bands, and
PKS 1610$-$771 was observed every 45 minutes for antenna gain calibration.

\begin{deluxetable*}{cccccc}
\tablecaption{ATCA Observation Parameters of the Snail \label{table:para}}
\tablewidth{0pt}
\tabletypesize{\small}
\tablehead{\colhead{Obs.\ Date} & \colhead{Array Config.} & \colhead{Center
Freq.} & \colhead{Usable Bandwidth} & \colhead{No.\ of} &
\colhead{Integration} \\
& & \colhead{(MHz)\tablenotemark{1}} & \colhead{(MHz)\tablenotemark{1}} &
\colhead{Channels\tablenotemark{1}} & \colhead{Time (hr)}}
\startdata
2000 Apr 17 & 1.5D & 1420, 2496 & 4, 104 & 1025, 13 & 9.1  \\
2000 Apr 29 & 750D & 1420, 2240 & 4, 104 & 1025, 13 & 9.9 \\
2008 Dec 09 & 750B & 4800, 8640 & 104, 104 & 13, 13 & 10.4 \\
2009 Feb 19 & EW352 & 4800, 8640 & 104, 104 & 13, 13 & 10.4 \\ 
\enddata
\tablenotetext{1}{Per center frequency, respectively.}         
\end{deluxetable*}
\begin{deluxetable}{cccc}
\tablecaption{Flux Densities and Spectral Indices of Different Parts of the
Snail \label{table:flux}}
\tablewidth{0pt}
\tabletypesize{\small}
\tablehead{\colhead{$\lambda$ (cm)} & \colhead{Whole PWN} & \colhead{The Body}
& \colhead{The Head}}
\startdata
36 & $2.3 \pm 0.2\,{\rm Jy}$ & $2.1 \pm 0.1\,{\rm Jy}$ & $0.18 \pm
0.02\,{\rm Jy}$ \\
20 & $2.1 \pm 0.4\,{\rm Jy}$ & $1.8 \pm 0.4\,{\rm Jy}$ & $0.30 \pm
0.06\,{\rm Jy}$ \\
13 & $1.8 \pm 0.3\,{\rm Jy}$ & $1.5 \pm 0.2\,{\rm Jy}$ & $0.23 \pm
0.03\,{\rm Jy}$ \\
6 & $1.5 \pm 0.2\,{\rm Jy}$ & $1.3 \pm 0.1\,{\rm Jy}$ & $0.16 \pm 0.02\,{\rm
Jy}$ \\
3 & $0.94 \pm 0.06\,{\rm Jy}$ & $0.83 \pm 0.06\,{\rm Jy}$ & $0.11 \pm
0.01\,{\rm Jy}$ \\
\hline
$\alpha$ ($S_\nu \propto \nu^\alpha$) & $-0.3 \pm 0.1$\tablenotemark{a} &
$-0.3 \pm 0.1$\tablenotemark{a} & $-0.6 \pm 0.1$\tablenotemark{b}
\enddata
\tablenotetext{a}{Best fit to 36, 20, 13, and 6\,cm.}
\tablenotetext{b}{Best fit to 20, 13, 6, and 3\,cm.}
\end{deluxetable}

We used the MIRIAD package \citep{miriad} for all of the data reduction. First,
edge channels and channels known to be affected by radio frequency interference
were discarded. We then examined the data and flagged bad data points during
periods of poor atmospheric stability. Next, bandpass, gain, polarization, and
flux calibration solutions were determined. In order to obtain a uniform
\textit{u-v} coverage, we excluded data from the 6\,km baseline while forming
radio maps. We formed radio images for the four bands, adopting the
multifrequency synthesis technique \citep{mfs} which can further improve the
\textit{u-v} coverage. In particular, the 20\,cm continuum image was formed by
extracting line-free channels from the data. Uniform weighting was used for the
20, 13, and 6\,cm images to reduce sidelobes and to improve the spatial
resolution.  For the 3\,cm image, due to the lower signal-to-noise (S/N) ratio
in this band, we used a weighting parameter of \texttt{robust} $ = 0$
\citep{robust} to optimize the balance between the noise level and spatial
resolution. We then applied a maximum entropy algorithm \citep{maxen} to
deconvolve all of the dirty maps. For 20 and 13\,cm, the task \texttt{MAXEN}
was employed; for 6 and 3\,cm, we used the task \texttt{PMOSMEM} \citep{mosmem}
to deconvolve the Stokes \textit I, \textit Q, and \textit U maps
simultaneously.  After that, we convolved the resultant models with synthesized
beams of FWHM $26^{\prime\prime} \times 19^{\prime\prime}$ for 20\,cm,
$16^{\prime\prime} \times 11^{\prime\prime}$ for 13\,cm, $15^{\prime\prime}
\times 13^{\prime\prime}$ for 6\,cm, and $10^{\prime\prime} \times
8^{\prime\prime}$ for 3\,cm. The final maps have an rms noise of $0.44\,{\rm
mJy\,beam^{-1}}$ at 20\,cm, $0.085\,{\rm mJy\,beam^{-1}}$ at 13\,cm,
$0.15\,{\rm mJy\,beam^{-1}}$ at 6\,cm, and $0.11\,{\rm mJy\,beam^{-1}}$ at
3\,cm. These all agree well with the theoretical values. We also formed images
with visibilities from the 6\,km baselines only in an attempt to identify point
sources. Identical data reduction and imaging procedures as above were used,
except that natural weighting was used to form the images in order to reduce
the noise.

\begin{figure}[ht]
\epsscale{1.1}
\plotone{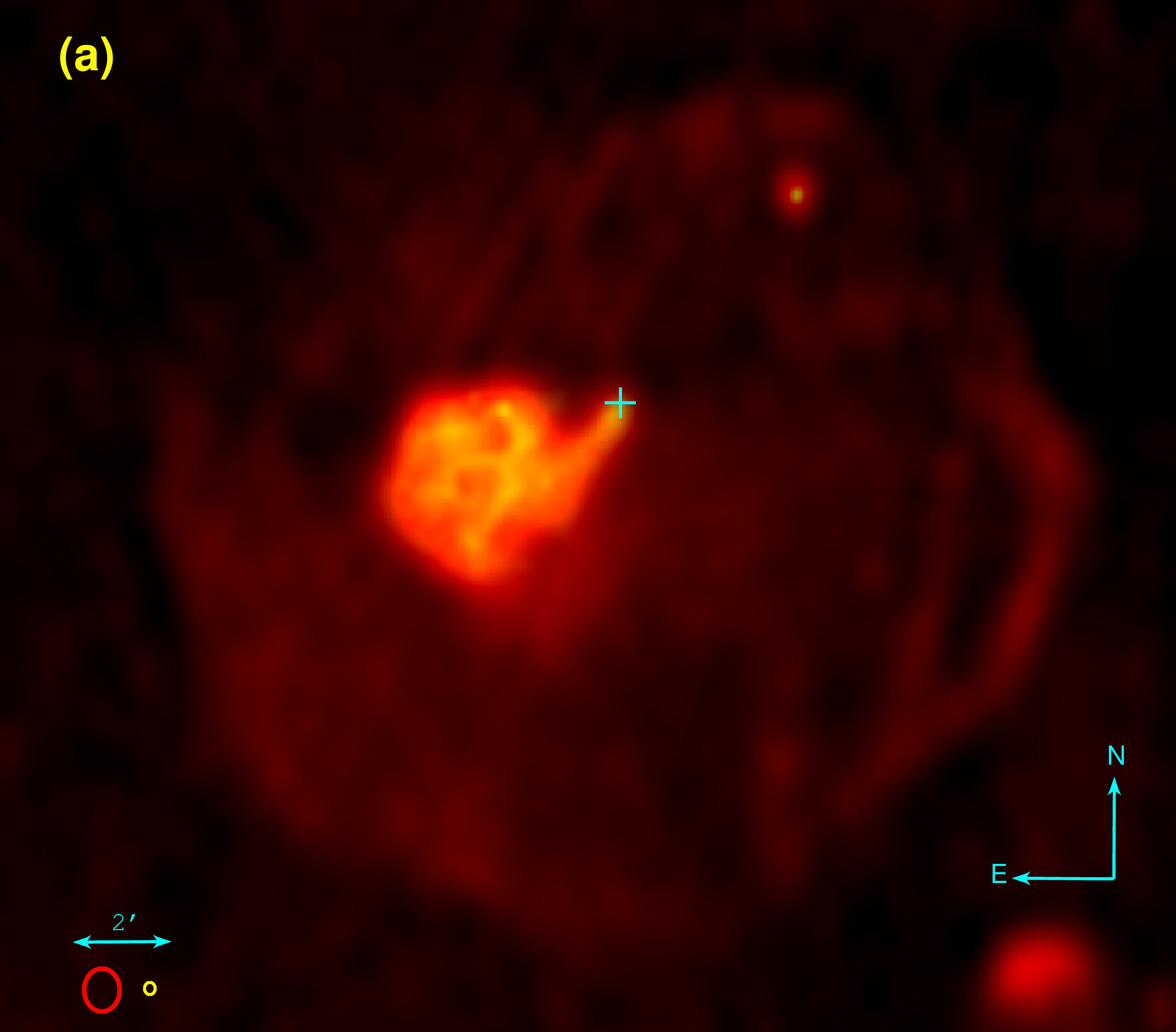}
\plotone{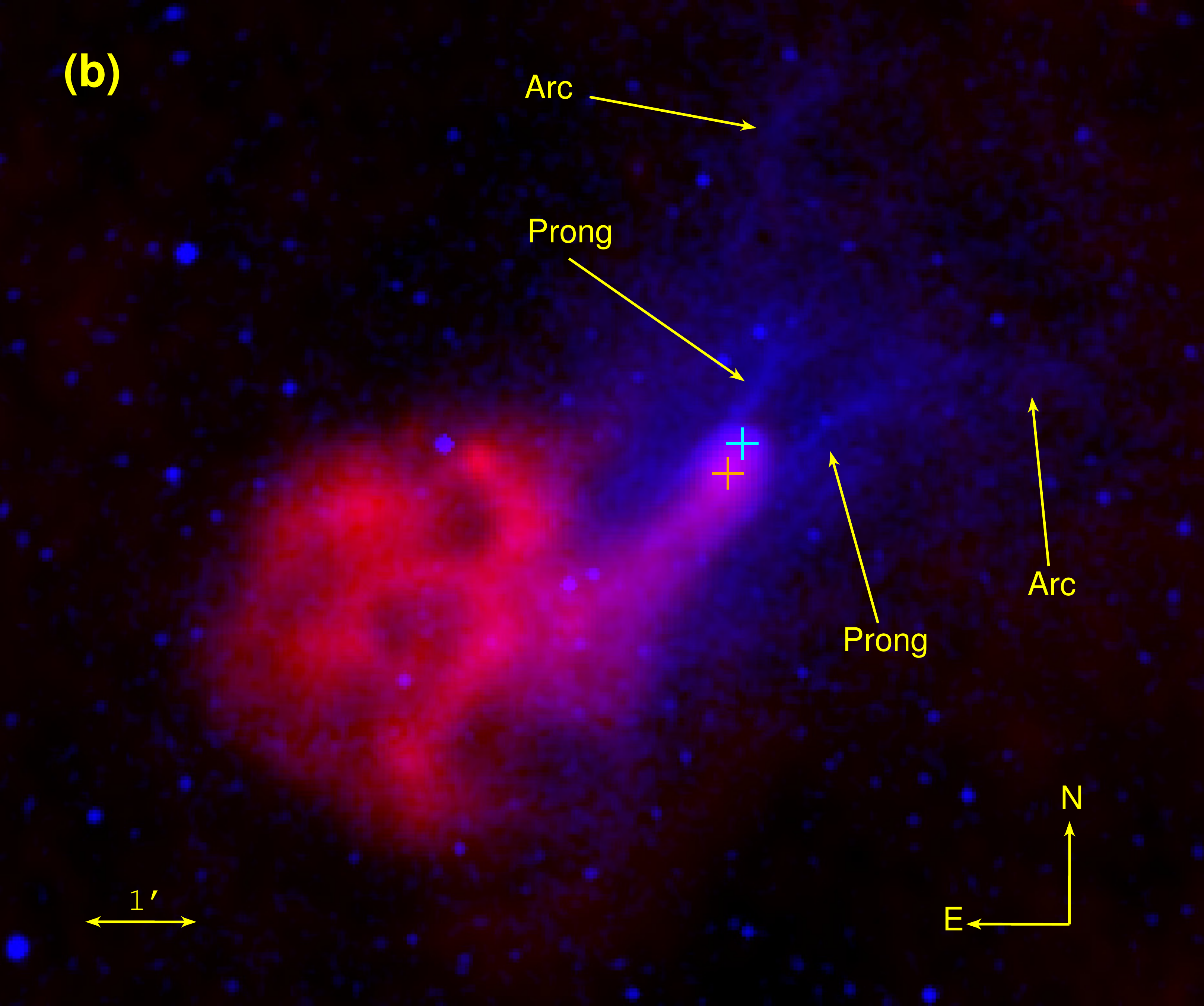}
\caption{(a) Composite radio image of SNR G327.1$-$1.1. The MOST 36\,cm
intensity map \citep{whiteoak96} is shown in red, illustrating the SNR shell
and the PWN. The ATCA 6\,cm map is shown in yellow to highlight the detailed
structures in the PWN. The radio beams at 36 and 6\,cm are shown in the lower
left corner as red and yellow ellipses, respectively. The cyan cross marks the
position of the X-ray point source reported by \cite{temim09}, which is
believed to be an associated neutron star. Note that the ATCA observations at
6\,cm are only sensitive to a scale smaller than $\sim 7^\prime$, and therefore
are not expected to be sensitive to the SNR shell. (b) Comparison between the
radio and X-ray emission of the Snail. The ATCA 6\,cm radio image is shown in
red and the \textit{Chandra} ACIS image in $0.5$--$7\,{\rm keV}$
\citep{temim15} is shown in blue. The latter is smoothed to a resolution of
$4^{\prime\prime}$.  The X-ray point source is marked by the cyan cross and the
radio peak is marked by the orange cross. Note that the sizes of the crosses do
not represent the respective astrometric uncertainties. \label{fig:g327}}
\end{figure}
\begin{figure*}[ht]
\epsscale{0.95}
\plotone{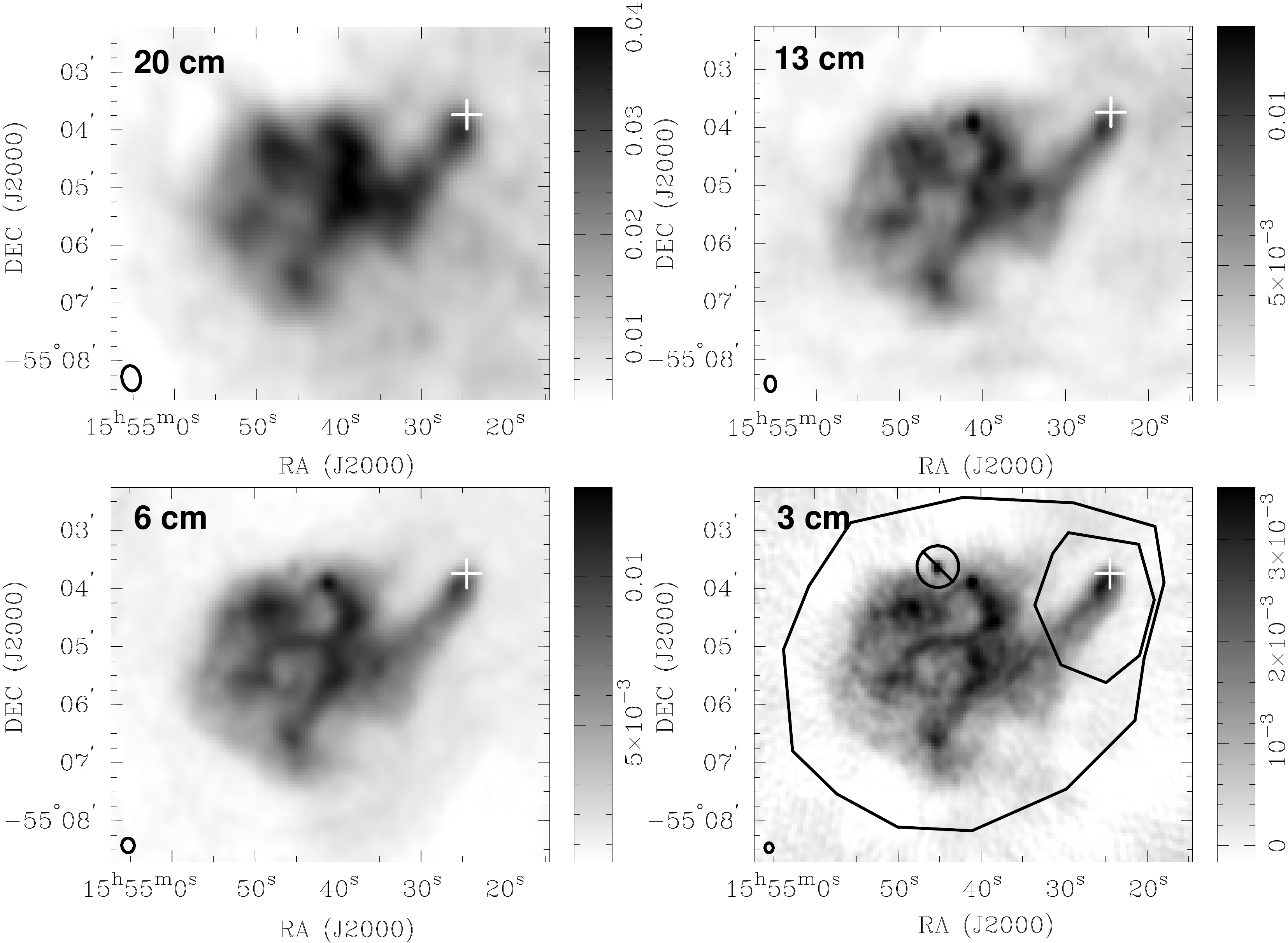}
\caption{Radio intensity maps of the Snail at 20, 13, 6, and 3\,cm. The
restoring beams are shown in the lower left corner of each maps. The grayscales
are linear and the scale bars have units of ${\rm Jy\,beam}^{-1}$. The crosses
mark the position of the X-ray point source. The regions adopted for flux
density measurements are also shown in the 3\,cm map. The large polygon
indicates the extraction region for flux density measurement of the entire PWN,
and the smaller one is for the head only. The body region is that of the entire
PWN excluding the head. The crossed out circle centered on the eastern point
source is an exclusion region.\label{fig:int}} 
\end{figure*}

For polarimetry, we formed a new set of low-resolution 3\,cm images matching
that at 6\,cm in order to make a direct comparison between the two bands and
generate a rotation measure (RM) map (see Section~\ref{sec:rm}). A Gaussian
taper has been applied to the visibility data by setting the parameter of
\texttt{FWHM} $= 15$, $13$ in the task \texttt{INVERT}. We then followed the
same imaging and cleaning procedures as outlined above, except we convolved the
output model of the task \texttt{PMOSMEM} with a Gaussian beam of identical
size as the 6\,cm beam (${\rm FWHM} = 15^{\prime\prime} \times
13^{\prime\prime}$) in the task \texttt{RESTOR}. We obtained polarized
intensity and polarization position angle (PA) maps from the Stokes \textit Q
and \textit U maps at the two bands. The Ricean bias was corrected for when the
polarized intensity was computed \citep{bias}. We blanked out areas where
either the polarized intensity has S/N $<$ 5 or the total intensity has S/N $<$
12.  Note that polarization was not measured at 20\,cm and the 13\,cm data are
not useful, since the Stokes \textit Q and \textit U maps are corrupted by
severe polarization leakage from a nearby bright H~{\sc ii} region G327.3$-$0.6.
This is due to a flaw in the feedhorn design of the old ATCA
receiver\footnote{\url{http://www.atnf.csiro.au/computing/at\_bugs.html\#Bug\_19}.
We attempted to correct for the polarization leakage with the task
\texttt{OFFAXIS} but had no success.}.

For the H~{\sc i} data, we used the task \texttt{UVLIN} to separate line
emission from continuum emission within the 20\,cm data. We then smoothed the
line spectrum to a resolution of $3.3\,{\rm km\,s^{-1}}$ in velocity. To filter
out large-scale structures, only data with \textit{u-v} distances shorter than
$1\,{\rm k}\lambda$ were selected to form the line images. We found that the
source is too weak to determine any H~{\sc i} absorption.

\section{RESULTS} \label{sec:results}
\subsection{PWN Morphology} \label{sec:morph}
A composite radio image of SNR G327.1$-$1.1 at 36\,cm
\citep[MOST;][]{whiteoak96} and 6\,cm (ATCA) is shown in
Figure~\ref{fig:g327}a, featuring both the SNR shell and the PWN. Radio
intensity maps of the PWN at 20, 13, 6, and 3\,cm (all from ATCA) are shown in
Figure~\ref{fig:int}. The nebula exhibits similar morphology over the observed
wavelengths. It has a main circular structure with a diameter of $\sim
4^\prime$ and a finger-like structure of length $\sim 1\farcm5$ protruding
toward the northwest. Together with the two X-ray prongs ahead of the radio
finger, the whole PWN resembles a snail. We thus refer to the entire PWN as
``the Snail'': the circular structure is the ``body'' and the protrusion is the
\begin{figure}[ht]
\epsscale{1.15}
\plotone{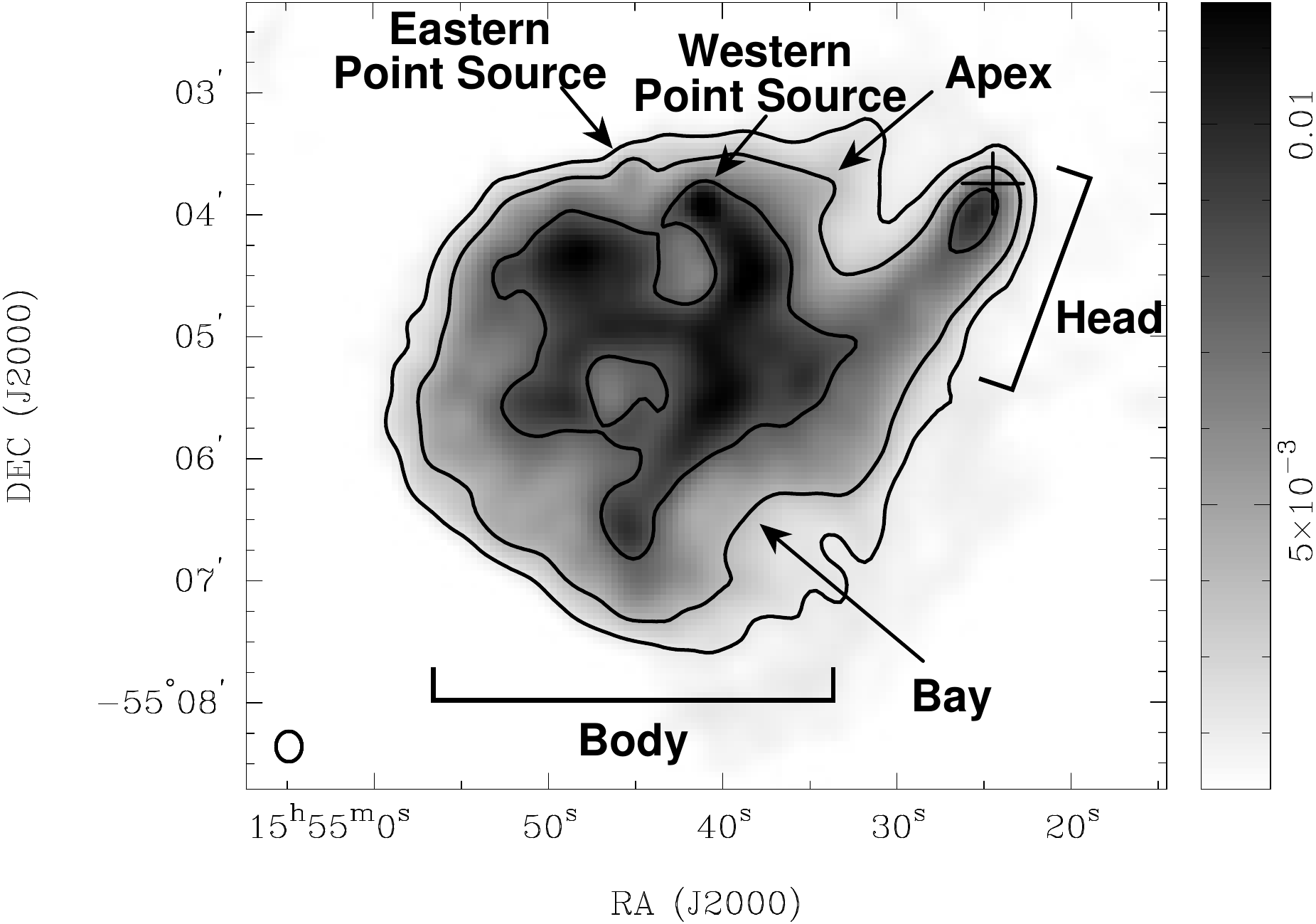}
\caption{Structure of the Snail. Both the grayscale and contours are from
the 6\,cm intensity map, with contour levels set at $2.2$, $4$, and $8\,{\rm
mJy\,beam^{-1}}$. The cross marks the position of the X-ray point source. The
restoring beam at 6\,cm is shown in the lower left corner. The eastern point
source can be seen more clearly in the 3\,cm image in Figure~\ref{fig:int}.
\label{fig:structure}}
\end{figure}
``head.'' A more detailed look at the body revealed two features --- ``the
bay'' to the southwest and ``the apex'' to the northwest (see
Figure~\ref{fig:structure}).  The bay is an indented area to the body, while
the apex is a sharp corner at the body's boundary. Faint emission is detected
outside of the bay at 20 and 6\,cm (about $13.8\,{\rm mJy\,beam^{-1}}$ and
$2.5\,{\rm mJy\,beam^{-1}}$, respectively), and was also found in the MOST
image \citep{sun99}. The emission is not obvious in the 13 and 3\,cm images.
This could be attributed to the contamination of the 13\,cm image by the SNR,
while at 3\,cm it could be too faint to be detected, or could have been
resolved out because of insufficient \textit{u-v} coverage (see
Section~\ref{sec:spec}). On a larger scale, hints of an SNR shell of diameter
$\sim 17^\prime$ are seen in the 20 and 13\,cm images (not shown in the field
of view of Figure~\ref{fig:int}).  Since the shortest baseline of the 6 and
3\,cm observations is 31\,m, which translates to angular scales of about
$7^\prime$ and $4^\prime$, respectively, we do not expect these observations to
be sensitive to the SNR shell.

The 13, 6, and 3\,cm images clearly reveal small-scale structure within the
body. We discovered filaments running through the PWN with an unresolved width
by our observations. These filaments seem to form loop-like structures that are
typically $\sim 1^\prime$ diameter. There are two point sources near the
northern edge of the body. The eastern one is located at ($15^{\rm h}54^{\rm
m}45^{\rm s}$, $-55^\circ 03^\prime 39^{\prime\prime}$; J2000.0) and the
western one is at ($15^{\rm h}54^{\rm m}41^{\rm s}$, $-55^\circ 03^\prime
55^{\prime\prime}$; J2000.0). These are marked in Figure~\ref{fig:structure}.
Both point sources are unresolved in the 6\,km baseline images, and therefore
are less than $1^{\prime\prime}$ in extent.

The head has an extent of about $1\farcm5$ long and $0\farcm4$ wide, and shows
uniform surface brightness ($\sim 7\,{\rm mJy\,beam^{-1}}$ at 6\,cm, and $\sim
2\,{\rm mJy\,beam^{-1}}$ at 3\,cm), except near the tip ($15^{\rm h} 54^{\rm m}
25.4^{\rm s}$, $-55^\circ 04^\prime 01.3^{\prime\prime}$; J2000.0) where it
shows a peak with $\sim 9\,{\rm mJy\,beam^{-1}}$ at 6\,cm and $\sim 3\,{\rm
mJy\,beam^{-1}}$ at 3\,cm. The radio peak is extended, with a size of $\sim
34^{\prime\prime} \times 18^{\prime\prime}$ elongated along the head, and is
offset from the X-ray point source by $\sim 18^{\prime\prime}$ to the
southeast. No radio counterparts of the X-ray prongs and arc outside of the
head are detected. Careful examination of the 6\,km baseline images shows no
counterparts of the X-ray point source with a $3\sigma$ upper limit of
$1.1\,{\rm mJy}$ at 6\,cm.

\begin{figure}[ht]
\epsscale{1.04}
\plotone{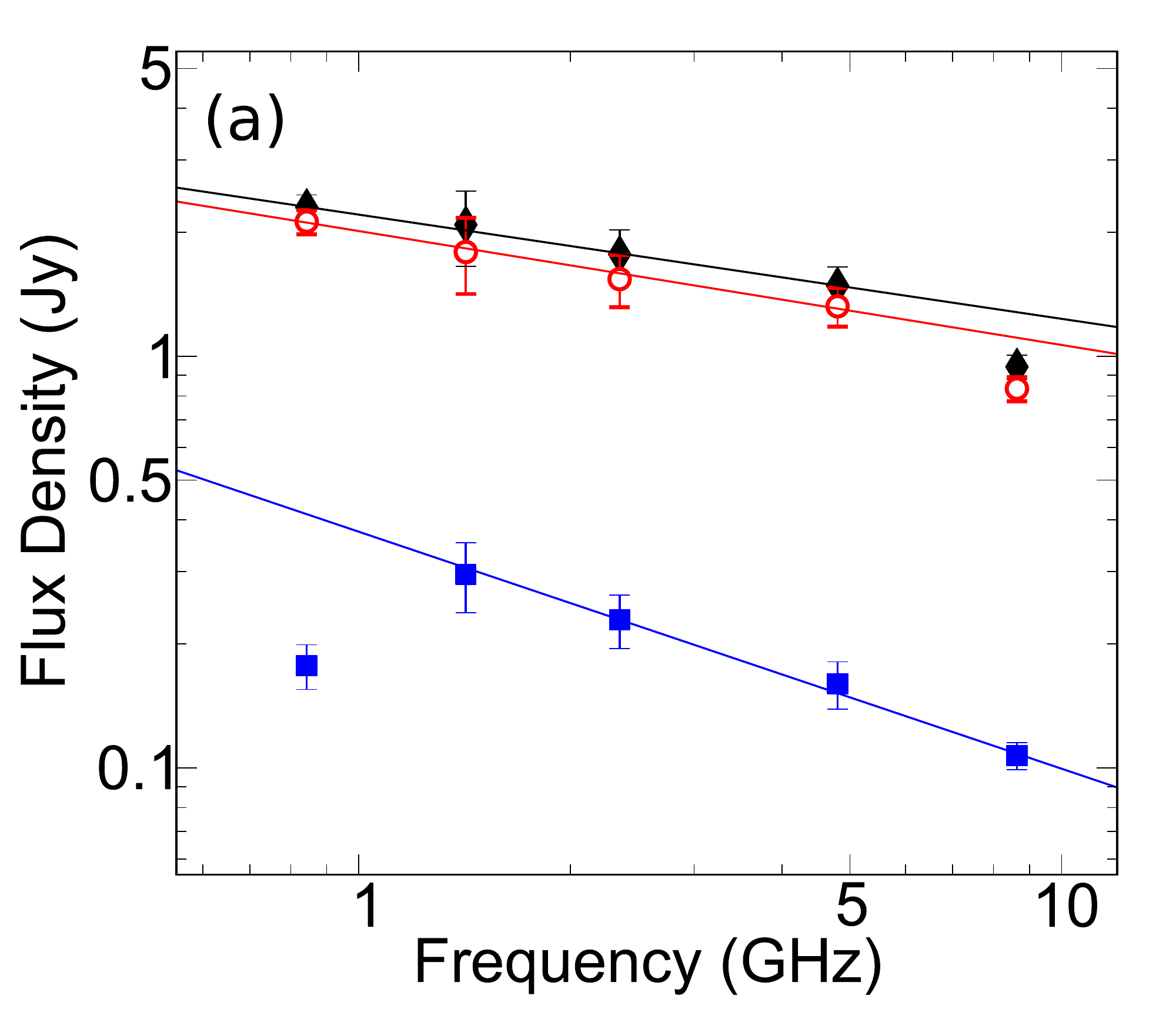}
\plotone{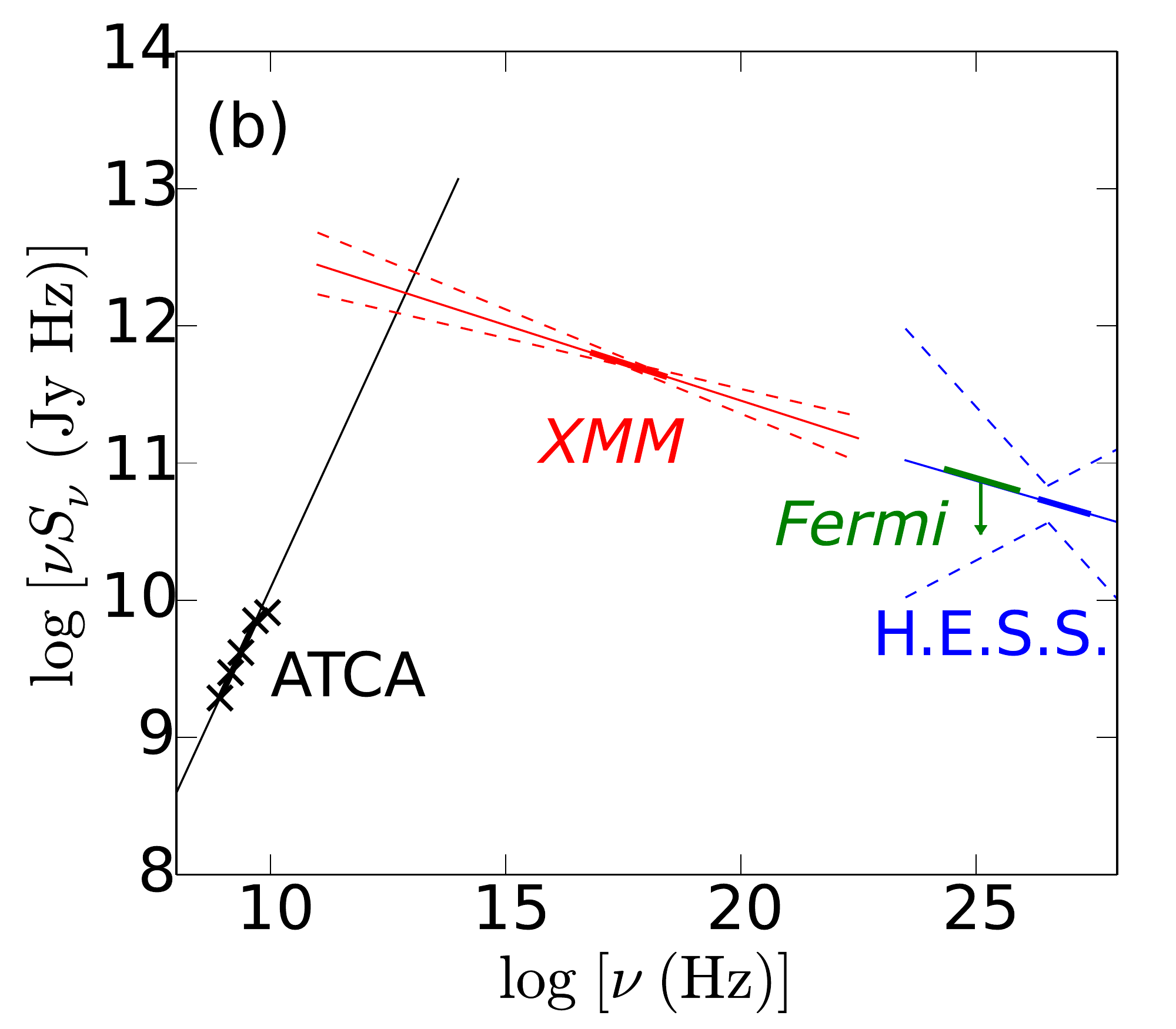}
\caption{(a) Radio spectra of the Snail. Flux densities of the entire PWN, the
body, and the head are marked by black closed diamonds, red open circles, and
blue closed squares, respectively. The best fits to the three spectra give
spectral indices of $\alpha = -0.3 \pm 0.1$, $-0.3 \pm 0.1$, and $-0.6 \pm
0.1$, respectively (with $S_\nu \propto \nu^\alpha$). (b) Spectral energy
distribution of the Snail from the radio to TeV band
\citep{temim09,acero12,acero13}. The best-fit spectra within the respective
observed bands are shown as thick solid lines and the extrapolations are shown
as thin solid lines. The uncertainties of the fitted spectra are shown as
dashed lines. \label{fig:radiospec}} 
\end{figure}

\subsection{Radio Spectrum} \label{sec:spec}
To measure the PWN flux density, we employed a source region enclosing both the
body and the head as shown in Figure~\ref{fig:int}. The eastern point source is
believed to be a background source while the western point source might be a
compact structure within the PWN (see Section~\ref{sec:ewsrc}). Therefore, we
defined a circular exclusion region of $43^{\prime\prime}$ diameter centered at
the former. We tried different background regions within the SNR shell, and the
difference gives us a handle on the systematic uncertainty. The results are
shown in Table~\ref{table:flux}. The flux densities of the PWN are $2.3 \pm
0.2\,{\rm Jy}$, $2.1 \pm 0.4\,{\rm Jy}$, $1.8 \pm 0.3\,{\rm Jy}$, $1.5 \pm
0.2\,{\rm Jy}$, and $0.94 \pm 0.06\,{\rm Jy}$ at 36, 20, 13, 6, and 3\,cm,
respectively. The uncertainties are dominated by the systematic uncertainties
associated with background subtraction, as the statistical uncertainties are
relatively small (a few mJy). We note that the value we find at 36\,cm
(2.3\,Jy) is slightly larger than that (2.0\,Jy) reported by \cite{whiteoak96}.
This could be attributed to our different choice of regions.

The radio spectrum is shown in Figure~\ref{fig:radiospec}. Fitting the flux
densities at 36, 20, 13, and 6\,cm with a power law ($S_\nu \propto
\nu^\alpha$) gives a spectral index of $\alpha = -0.3 \pm 0.1$, and
extrapolation to 3\,cm suggests a flux density of $1.3\,{\rm Jy}$, which is
almost $40\%$ higher than the observed value of $0.94\,{\rm Jy}$. If we
directly join the 6 and 3\,cm data points, then we obtain $\alpha = -0.8 \pm
0.2$ between the two bands. The shortest \textit{u-v} spacing at 3\,cm is about
$0.8\,{\rm k}\lambda$, corresponding to an angular scale of $4\farcm3$. This
sets a rough upper limit on the size of structures that our observations are
sensitive to.  In order to see if the missing flux problem is significant, we
formed a 6\,cm radio image only with \textit{u-v} distance larger than
$0.8\,{\rm k}\lambda$, using identical procedure as described in
Section~\ref{sec:obs}. This results in about $20\%$ loss in flux density of the
PWN, suggesting that the 3\,cm observations may not be sensitive to the
larger-scale emission of the PWN.

We also determined the flux densities of the body and the head of the Snail
separately, with the results also listed in Table~\ref{table:flux}. The flux
densities and fitted spectra are shown in Figure~\ref{fig:radiospec}. For the
body, the flux densities are $2.1 \pm 0.1\,{\rm Jy}$, $1.8 \pm 0.4\,{\rm Jy}$,
$1.5 \pm 0.2\,{\rm Jy}$, $1.3 \pm 0.1\,{\rm Jy}$, and $0.83 \pm 0.06\,{\rm Jy}$
at 36, 20, 13, 6, and 3\,cm, respectively, and fitting the flux densities
between 36 and 6\,cm with a power-law spectrum gives a spectral index of
$\alpha = -0.3 \pm 0.1$. For the head, the flux densities are $0.18 \pm
0.02\,{\rm Jy}$, $0.30 \pm 0.06\,{\rm Jy}$, $0.23 \pm 0.03\,{\rm Jy}$, $0.16
\pm 0.02\,{\rm Jy}$, and $0.11 \pm 0.01\,{\rm Jy}$ at 36, 20, 13, 6, and 3\,cm,
respectively. The spectrum of the head appears to be peculiar, as the observed
flux density is lower at 36\,cm than that at 20\,cm. We believe that the flux
density measurement of the head could be affected by faint sidelobes from
G327.3$-$0.6 at 36\,cm. We fitted the flux densities of the head between 20 and
3\,cm, as the size of the head is only about $1\farcm5$ long and $0\farcm4$
wide, and should be well sampled by our 3\,cm observations. The resulting
power-law spectrum has a spectral index of $\alpha = -0.6 \pm 0.1$.

\begin{figure}[ht]
\epsscale{1.1}
\plotone{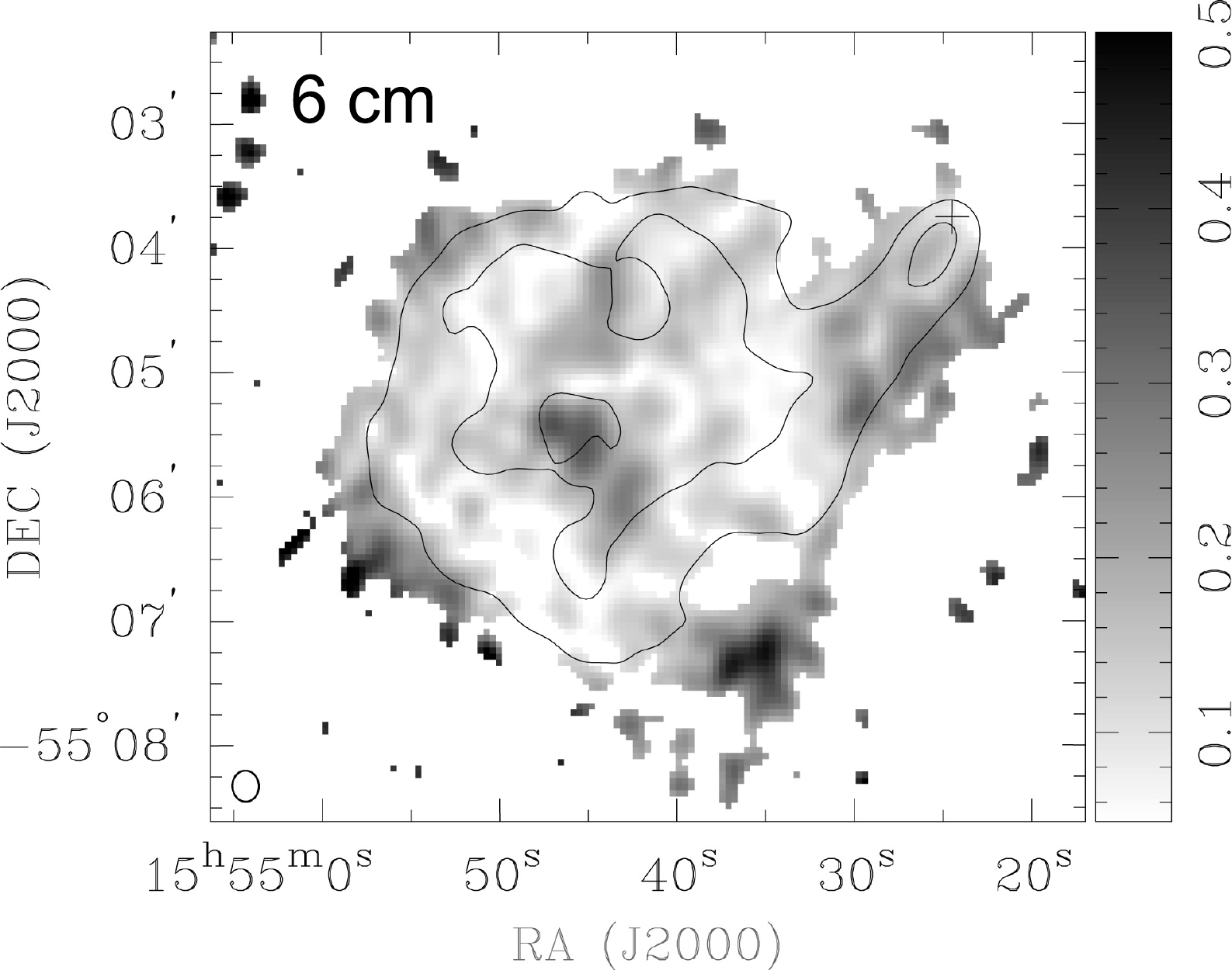}
\plotone{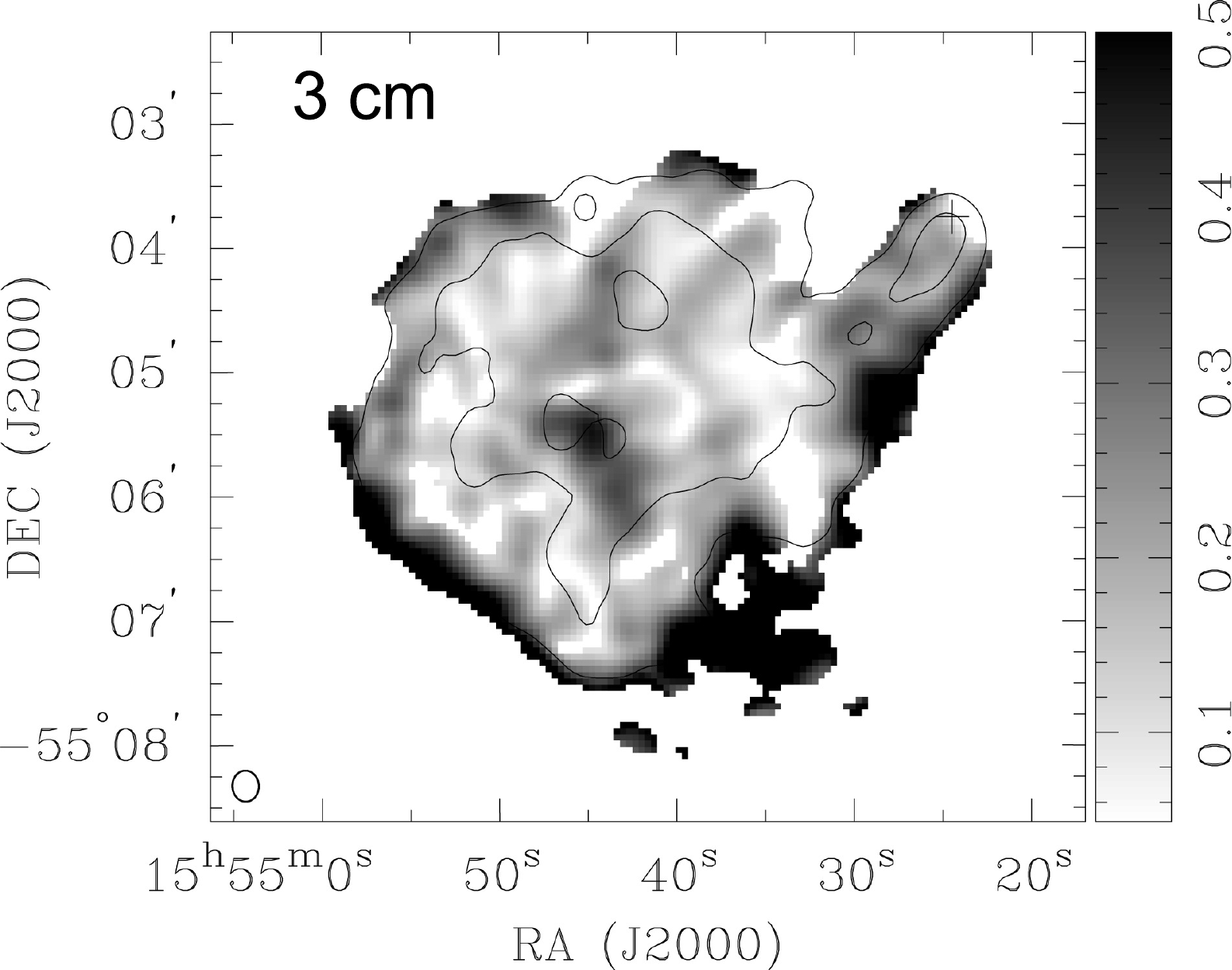}
\caption{Fractional polarization of the Snail at 6 and 3\,cm. The contours are
from the total intensity map of $4$ and $8\,{\rm mJy\,beam}^{-1}$ at 6\,cm,
and $1.5$ and $4.5\,{\rm mJy\,beam}^{-1}$ at 3\,cm. FWHM of the beams are
shown in the lower left. The crosses mark the position of the X-ray point
source. Note that the polarization at the edge is likely overestimated
(see Section~\ref{sec:pol}). \label{fig:pm}}
\end{figure}

The flux densities of the eastern point source are found to be $\sim 2.0\,{\rm
mJy}$ at 6\,cm and $\sim 3.3\,{\rm mJy}$ at 3\,cm, while those for the western
point source are $\sim 11\,{\rm mJy}$ at 6\,cm and $\sim 7.5\,{\rm mJy}$ at
3\,cm. These are very rough estimates, as the diffuse emission from the PWN
precludes precise measurements. The results suggest an inverted spectrum with
$\alpha \sim +0.9$ for the eastern point source and $\alpha \sim -0.7$ for the
western point source.

\subsection{Polarization} \label{sec:pol}
Maps of the polarization fraction of the Snail at 6 and 3\,cm are shown in
Figure~\ref{fig:pm}. Overall, the PWN is highly linearly polarized. The typical
polarization fraction of the body is about $15\%$ and $20\%$ at 6 and 3\,cm,
respectively, while that along the head is correspondingly about $20\%$ and
$30\%$. At a smaller scale, we found a highly polarized core inside the body as
enclosed by an inner contour in Figure~\ref{fig:pm}, with polarization
fractions of about $30\%$ and $40\%$ at 6 and 3\,cm, respectively.  Near the
northern edge of the body, the eastern point source is unpolarized, while the
western point source is about $10\%$ polarized at both bands. Note that the
actual polarization fraction at 3\,cm could be lower due to the missing flux
problem as mentioned. Finally, the polarization fraction at the edge of the
body may be overestimated because we have used only a single RMS value for the
debiasing procedure.

\begin{figure}[ht]
\epsscale{1.1}
\plotone{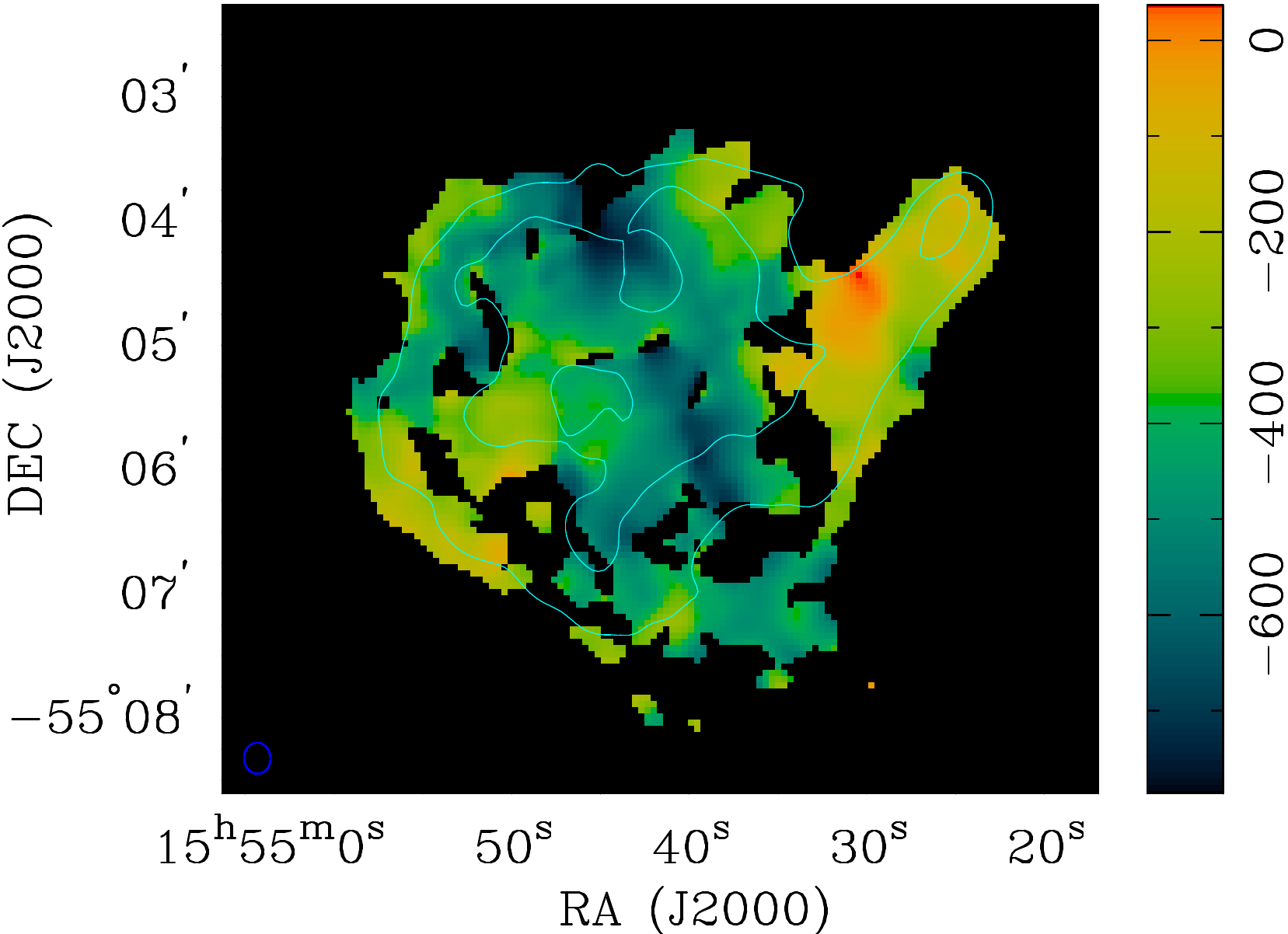}
\caption{RM map of the Snail. Contours are from the 6\,cm image of the PWN at
levels of $4$ and $8\,{\rm mJy\,beam}^{-1}$. The color map represents
the RM distribution of the Snail in units of $\rm{rad\,m}^{-2}$. The restoring
beam is shown at the lower left corner.} \label{fig:rm}
\end{figure}

\subsection{RM of the Snail} \label{sec:rm}
We generated an RM map by comparing between the polarization PA maps at 6 and
3\,cm: 
\begin{equation} 
{\rm RM} = \frac{\chi_6 - \chi_3}{\lambda_6^2 - \lambda_3^2}~{\rm ,}
\end{equation} 
where $\chi_6$ and $\chi_3$ are the observed PAs in the respective bands, and
$\lambda_6$ and $\lambda_3$ are the center wavelengths of the bands.  The
resulting RM map is shown in Figure~\ref{fig:rm}, and the RM is found to vary
smoothly between $-800$ and $0\,{\rm rad\,m^{-2}}$ across the PWN with a
typical uncertainty of $25\,{\rm rad\,m^{-2}}$. The average RM value is about
$-380\,{\rm rad\,m^{-2}}$.

\begin{figure*}[ht]
\epsscale{0.99}
\plotone{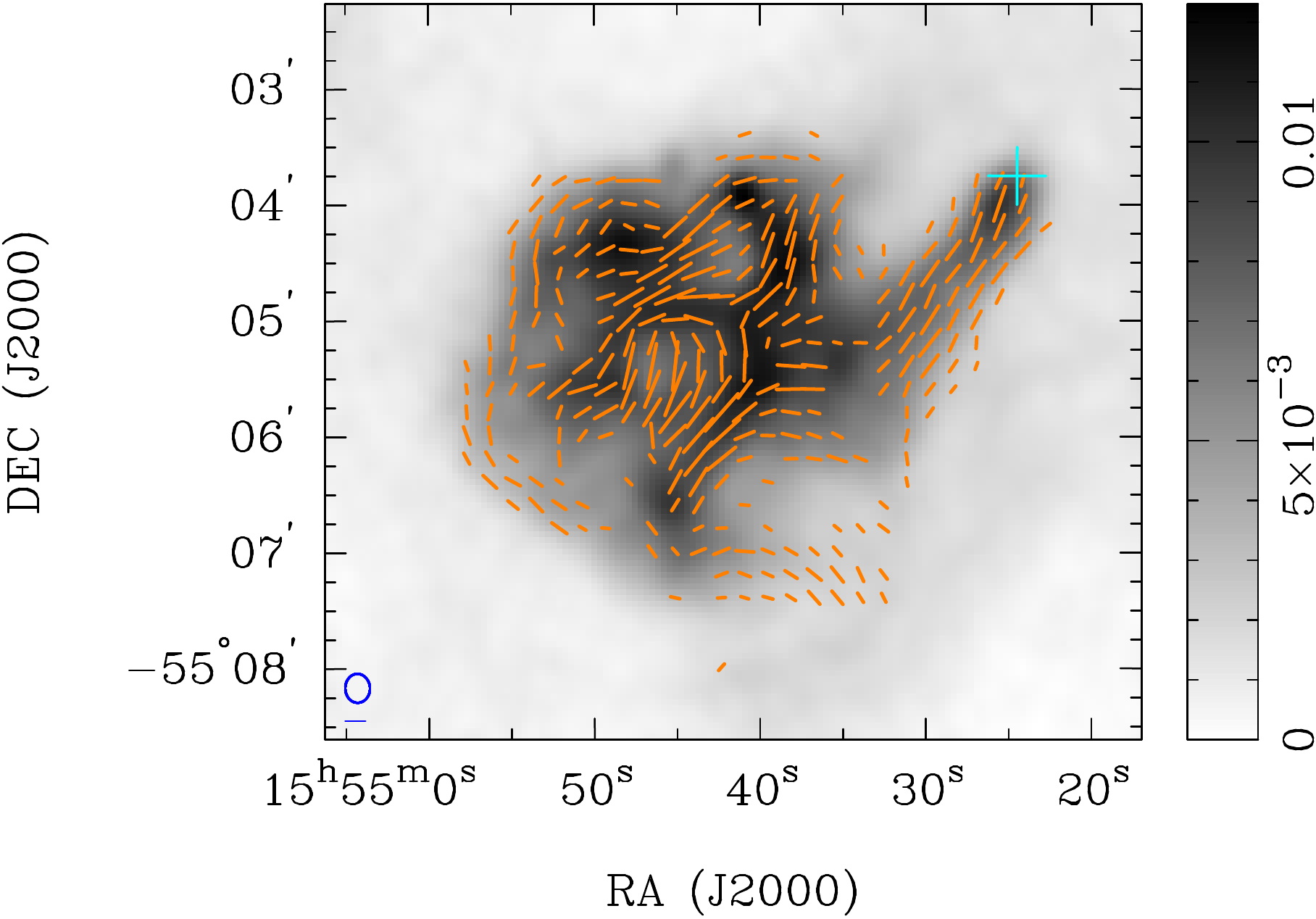}
\caption{Intrinsic magnetic field configuration of the Snail. The direction of
the vectors shows the orientation of the magnetic field corrected for Faraday
rotation, and the length is proportional to the polarized intensity with the
scale bar at the lower left representing $1\,{\rm mJy\,beam}^{-1}$. Typical
uncertainty in PA is about $3^\circ$. The grayscale shows the radio intensity
at 6\,cm with the scale bar in units of ${\rm Jy\,beam}^{-1}$. FWHM of the beam
is shown by the ellipse at the lower left. The cross marks the position of the
X-ray point source. \label{fig:bfield}} 
\end{figure*}

One problem of determining the RM this way is the so-called $n\pi$ ambiguity.
This ambiguity arises because an RM of larger magnitude can also be compatible
with the observed polarization vectors if it provides an addition of multiples
of $\pi$ radian to the relative rotation between the two bands. For example,
adding an extra RM of $1160\,{\rm rad\,m}^{-2}$ can rotate the polarization
vectors at 6 and 3\,cm by $\pi\,{\rm radian}$ relative to each other. In order
to eradicate this problem, we picked the first and last channels of the 6\,cm
data, which are at frequencies of $4848\,{\rm MHz}$ and $4752\,{\rm MHz}$,
respectively, and formed PA maps separately using procedures identical to those
of the full-band data. By applying the PA maps, we generated an RM map from the
two channels. Because the individual pixels of the RM map have large
uncertainty, we did not attempt to investigate the spatial distribution of the
RM from it.  Instead, we averaged the RM across the entire PWN to boost the S/N
ratio, and found an average RM for the Snail of $-380\,{\rm rad\,m}^{-2}$ from
the two channels, which is identical to the value obtained above. Since we
expect the rotation of the PA within the 6\,cm band to be small (less than
$\pi\,{\rm radian}$, otherwise the bandwidth depolarization would have been
significant), this rules out the $n\pi$ ambiguity in our measurements.

\subsection{Intrinsic Magnetic Field Orientation}
We employed the RM map obtained from Section~\ref{sec:rm} to derotate the
polarization vectors at 6 and 3\,cm in order to infer the intrinsic magnetic
field orientation of the PWN. Since the RM was derived from the PAs at 6 and
3\,cm only, the derotated vectors at the two bands have identical orientation.
The derotated vectors have a typical uncertainty of about $3^\circ$,
contributed by both the error in PA and in RM. Figure~\ref{fig:bfield} shows
the projected intrinsic magnetic field orientation of the Snail. Overall, the
field is highly ordered. The magnetic field at the head aligns with the PWN
elongation, while for the body it is tangential to the boundary, except at the
apex where it becomes radial. The field configuration of the interior of the
body is complex with the magnetic field generally following the filamentary
loops.

\section{DISCUSSION} \label{sec:discussion}
\subsection{PWN Structure and Magnetic Field Geometry}
\subsubsection{Overall PWN}
The Snail is believed to have first encountered the reverse shock about
$10\,{\rm kyr}$ ago, as suggested by HD simulations \citep{temim15}. This could
have resulted in a turbulent environment in the PWN and given rise to a tangled
magnetic field geometry. The strong polarization signal we found from the Snail
indicates the presence of a highly ordered magnetic field. This could reflect a
turbulence scale larger than the beam size of our observations.  In this case,
the beam depolarization will be insignificant and the observed magnetic field
geometry will appear ordered at small scales. For further investigation, MHD
efforts are required to fully understand the highly ordered magnetic field
found in the Snail.

\subsubsection{The Head}
The magnetic field in the head of the Snail shows good alignment with the
elongation of the structure (see Figure~\ref{fig:bfield}). This is consistent
with the evolutionary picture suggested by \cite{temim09}, assuming that the
magnetic field lines are frozen in the newly generated pulsar wind being pushed
to the southeast. Similar magnetic field geometry is also found in some
bow-shock PWNe, including the ``handle'' of the Frying Pan
\citep[G315.78$-$0.23;][]{ng12} and the tail of the Mouse
\citep[G359.23$-$0.82;][]{yusef05}. However, a distinctly different example is
the bow-shock system G319.9$-$0.7, which shows a helical magnetic field which
then aligns with the flow further downstream \citep{ng10}. The cause of such a
peculiar configuration is not well understood, though it may be related to the
Mach number of the pulsar with respect to its ambient medium or the relative
inclination of the spin axis and the elongation of the PWN
\citep[e.g.,][]{ng10,ng12}. As the head of the Snail is believed to be a
non-bow-shock system \citep[i.e.\ the Mach number of the neutron star $<
1$;][]{vds04,temim09,temim15}, it serves as an example showing that magnetic
field can align with subsonic comet-like PWNe.

We also compare the field geometry around the presumed pulsar with those of a
few evolved PWNe, including Vela X \citep{bock98,dodson03}, the Boomerang
\citep[G106.6$+$2.9;][]{kothes06}, and DA 495 \citep[G65.7$+$1.2;][]{kothes08}.
These PWNe could also have been crushed by the reverse shocks. The Boomerang
and DA 495 have diameters of about $1\,{\rm pc}$ and $4\,{\rm pc}$,
respectively. They are considerably smaller than the Snail, which has a length
of the head spanning $4\,{\rm pc}$ and the diameter of the body stretching
$10\,{\rm pc}$. The radio polarimetric observation of Vela X \citep{dodson03}
only covered the immediate vicinity ($\sim 0.4\,{\rm pc}$) around the Vela
pulsar. In the Boomerang and Vela X, toroidal magnetic fields can be observed,
while in DA 495 it was suggested to present a toroidal component superimposed
on the apparently dipolar field structure. The magnetic field structures in
these three PWNe could be driven by the pulsar itself
\citep{dodson03,kothes06,kothes08}, while in the head of the Snail the magnetic
field lines parallel to the nebulae elongation can be shaped by the passage of
the reverse shock. Note that the physical scale for which we have resolution of
in the Snail is very different from that of the three presented above. Future
high-resolution studies may potentially probe the pulsar-driven field near the
presumed pulsar of the Snail.

\subsubsection{Filaments in the Body}
The high spatial resolution ATCA images in Figure~\ref{fig:int} revealed the
filamentary structure inside the PWN. The filamentary loops generally align
with the magnetic field (see Figure~\ref{fig:bfield}), suggesting that they
could be magnetic loops similar to those observed in young PWNe 3C 58
\citep{slane04} and G54.1$+$0.3 \citep{lang10}. In the Snail, the filamentary
loops have a typical diameter of $\sim 1^\prime$, which corresponds to
$2.6\,{\rm pc}$ at a distance of $9\,{\rm kpc}$. The scale is a few times
larger than those in 3C 58 and G54.1$+$0.3, which were estimated to have
diameters of about $0.3$--$0.5\,{\rm pc}$ and $0.5$--$0.9\,{\rm pc}$,
respectively \citep{slane04,lang10}. It was suggested that kink instabilities
can be the origin of magnetic loops in PWNe --- the toroidal field near the
termination shock could be torn away and eventually become a magnetic loop as
observed \citep{slane04}. However, it is unclear if the interaction of the SNR
reverse shock could destroy these structures. In this case, the loops would be
formed through another mechanism.

\subsubsection{The Apex and the Bay}
We identified two structures of the Snail near the edge of the body referred
to as the apex and the bay. The former could be a protrusion from the body,
but it is difficult to confirm because its proximity to the head complicates
the interpretation of the morphology. The magnetic field vectors at the apex
show a radial instead of tangential orientation as seen in other parts of the
PWN boundary. Similar protrusions and chimney-like structures are seen in the
Crab Nebula \citep{bietenholz90} and in Kes 75 \citep{ng08}. The apex could be
formed by the leakage of PWN materials into the surrounding shocked SN ejecta,
similar to the chimney of the Crab Nebula. However, one difference between the
Crab Nebula and the body of the Snail is that the boundary magnetic field of
the Crab Nebula is primarily radial \citep{bietenholz90}, and so the physical
processes appear to be different.

The nature of the faint emission outside of the bay, which is outlined by the
outermost contour in Figure~\ref{fig:structure}, could be important for
understanding the evolution scenario of G327.1$-$1.1. It has been suggested as
part of the PWN \citep{sun99}, implying that the true boundary of the PWN
should lie outside the bay, such that the head would be pointing closer to the
center of the body as suggested by simulations \citep{temim15}, instead of
being tangential to its edge.

\subsubsection{Eastern and Western Point Sources} \label{sec:ewsrc}
As pointed out in Section~\ref{sec:morph}, our radio maps revealed two point
sources in the northern edge of the PWN. We found that the eastern one is
unpolarized (Figure~\ref{fig:pm}) and has an inverted spectrum of $\alpha \sim
+0.9$, which differs significantly from the typical value of $-0.3 \lesssim
\alpha \lesssim 0$ for PWNe. These points suggest that it is likely an
unrelated background source.

\begin{figure*}[ht!]
\epsscale{0.95}
\plotone{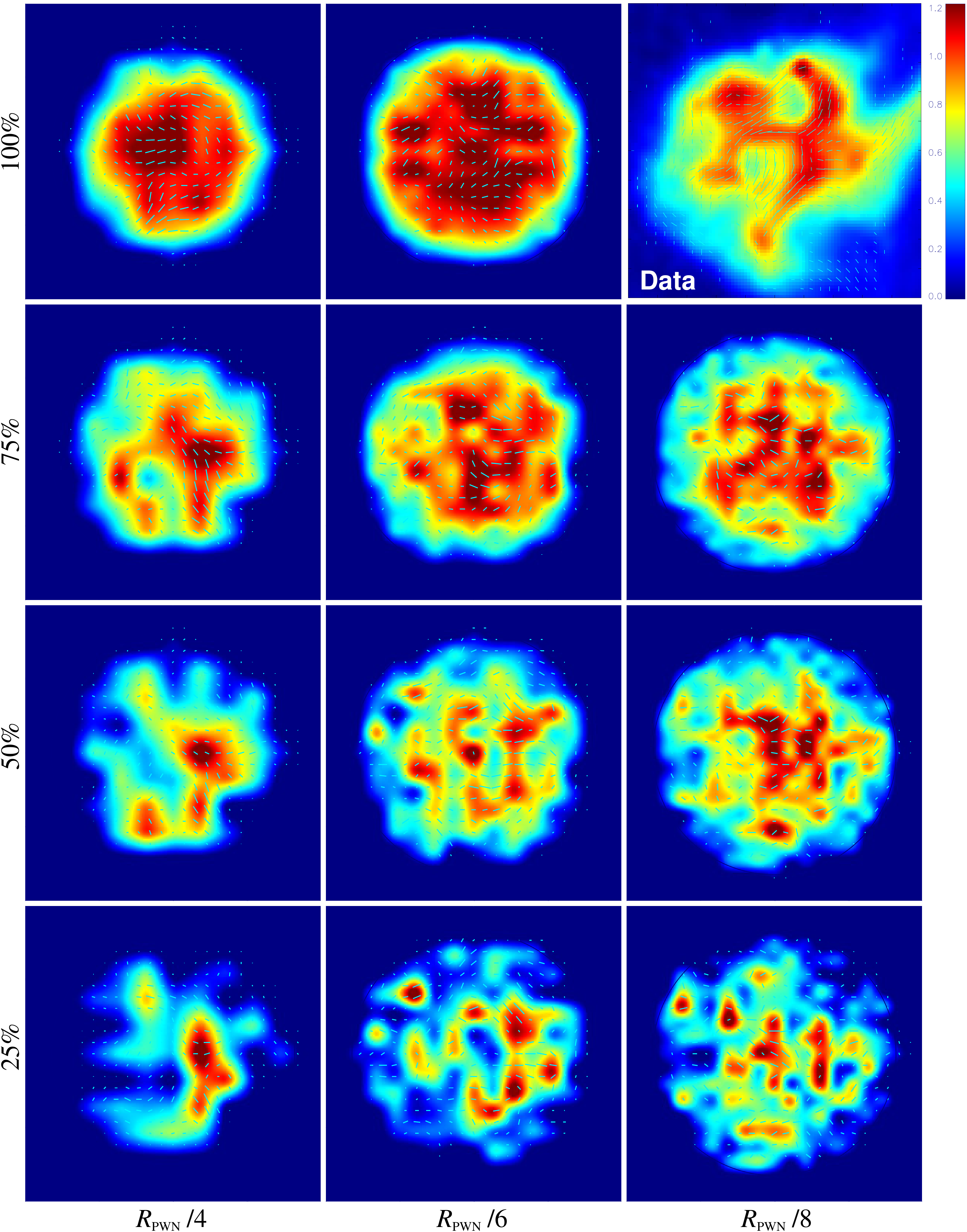}
\caption{Simulated intensity maps of a spherical PWN overlaid by the
polarization vectors compared with the data. The orientations of the vectors
show the nebular $B$-field directions and the lengths represent the polarized
intensity. Each model is comprised of patches of a certain size as indicated at
the bottom, with the magnetic field orientation randomly chosen for each of the
patches. The pulsar wind filling factor is also indicated to the left. The
observations of the Snail at 6\,cm are shown in the upper right panel with the
same color scale.\label{fig:mhd1}}
\end{figure*} 
\begin{figure*}[ht!]
\epsscale{0.95} 
\plotone{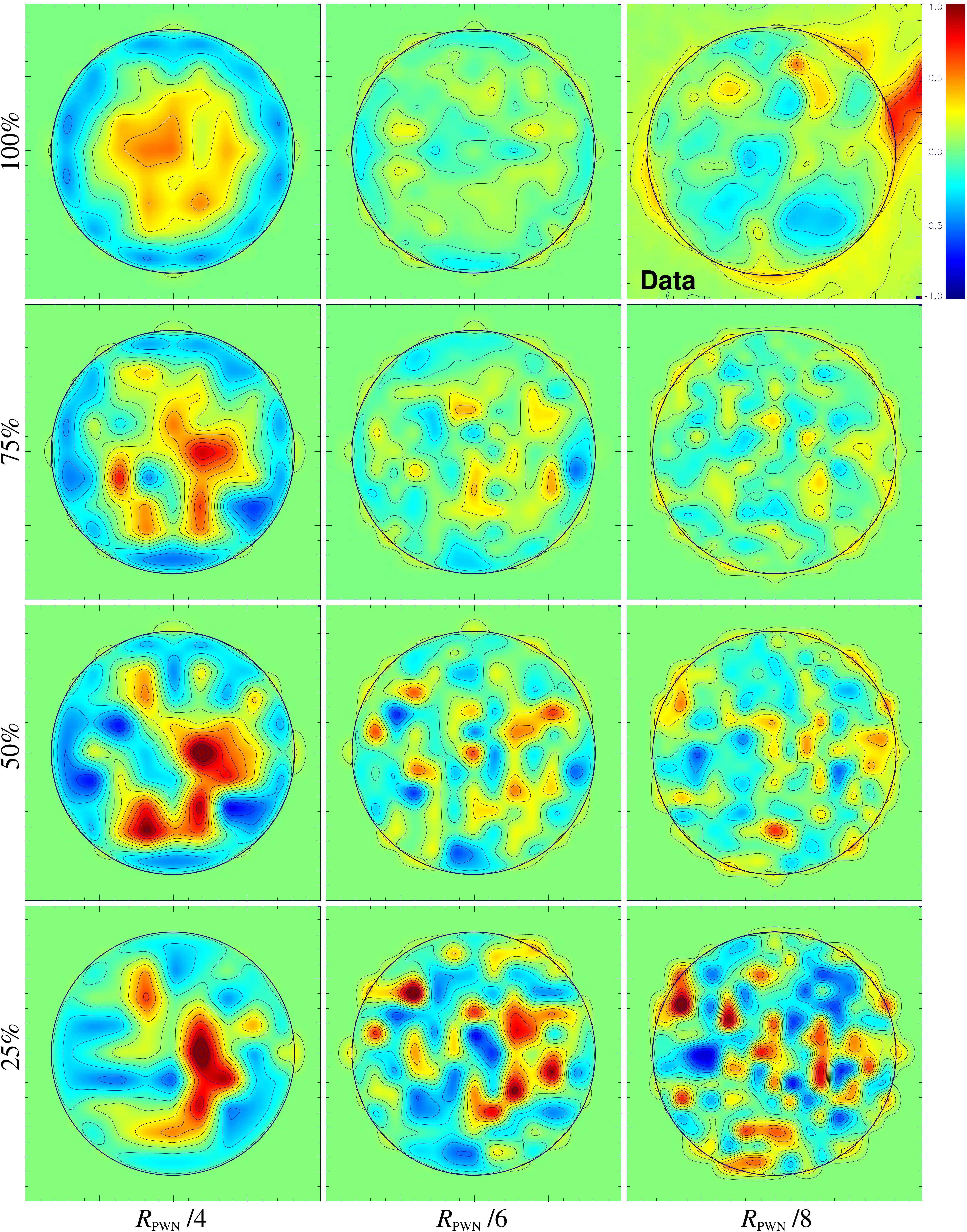} 
\caption{Intensity maps in Figure~\ref{fig:mhd1} subtracted by the mean
emissivity to show the fluctuation due to turbulence. The filling factors and
patch sizes are indicated at the left and bottom, respectively (see text for
details). The subtracted intensity map of the Snail is shown in the upper right
panel. \label{fig:mhd2}} 
\end{figure*} 
\begin{figure*}[ht!]
\epsscale{0.95} 
\plotone{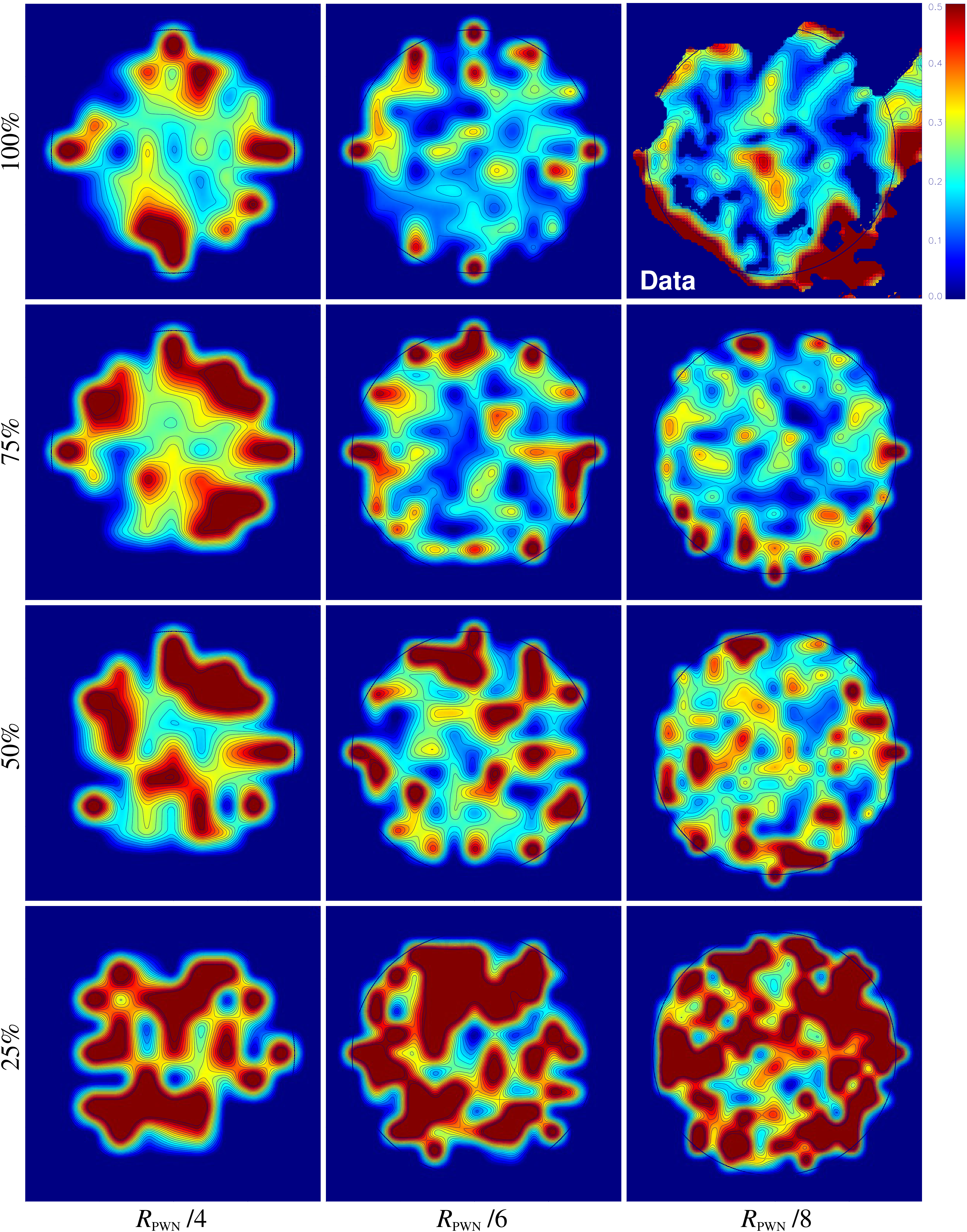} 
\caption{Simulated polarization maps compared with the data. The filling
factors and patch sizes are indicated at the left and bottom, respectively (see
text for details). The polarization map of the Snail at 6\,cm is shown in the
upper right panel. We note that the large polarization fraction near the edge
of the PWN is mostly due to noise, as the radio emission is weak.
\label{fig:mhd3}} 
\end{figure*}

On the other hand, the western source has a spectral index of $\alpha \sim
-0.7$ and is about $10\%$ polarized at 6 and 3\,cm. The polarized emission
allows us to determine the RM of the source and its value is consistent with
that of the surrounding PWN (see Figure~\ref{fig:rm}), suggesting a similar
distance. It also spatially coincides with the filaments in the PWN.
Therefore, we argue that it is associated with the Snail, possibly an
unresolved compact knot as part of the filaments.

\subsection{Simple Modeling of Turbulence in the Snail}
As mentioned earlier, the highly ordered magnetic field found in the Snail
could imply a large characteristic scale of the turbulence. To get a handle on
the scale in this case, we consider a toy model of the body of the Snail (i.e.\
excluding the head) as a non-evolving spherical nebula. This assumption is
based on the fact that we do not expect significant multipolar anisotropy in
the pressure exerted by the reverse shock, except for the dipolar component,
which will displace the PWN instead of deforming it. The volume of the nebula
is divided into patches of uniform size that represent the characteristic
turbulence scale in the PWN. The turbulent magnetic field is implemented in our
model by assigning a magnetic field with random orientation but uniform
strength to each of the patches. We argue that this assumption is justified, as
the magnetic field inferred from simulations \citep{temim15} is close to the
equipartition value we derived in Section~\ref{sec:bfield} below, and therefore
the fluctuations in the magnetic field strength should be small. Otherwise, the
fluctuation will rapidly fade away over a timescale of the order of the sound
crossing time (which is much shorter than the age of the system).  The electron
density is also assumed to be uniform for the same reason, and we assume a
power-law distribution with the index given by the measured radio spectrum. We
further allow for the possibility that a patch is filled with cold SNR material
instead of non-thermal pulsar wind particles. This mixing of SN ejecta in the
Snail was suggested based on X-ray observations \citep{temim15} and can also be
seen in other reverse shock crushed PWNe such as Vela X \citep{lamassa08}. The
probability that a patch is filled by synchrotron-emitting pulsar wind
particles is simply the filling factor in our model. If there is no significant
mixing of materials in the PWN, then the modeling results with a filling factor
of $100\%$ should fit the observations best, while a lower filling factor would
suggest substantial mixing.

We integrated the synchrotron emission along different lines of sight through
the volume of the nebula in all of the Stokes parameters to compute the radio
maps for comparisons with our observations. For this, we convolved the
simulation results with the radio beam. Note that the randomness introduced by
the filling factor and the magnetic field orientations make it possible to
exactly reproduce the observations if we run the model for a considerable
number of times (albeit that will be meaningless), and therefore it is more
sensible to only produce one set of maps per combination of parameters, and
then inspect by eye the scales of the structures, the number of features in the
PWN, and the contrast in radio intensity instead of looking for a perfect
match. This allows us to place a rough constraint on the filling factor and the
characteristic turbulence scale of the Snail. The simulated intensity and
fluctuation maps are shown in Figures \ref{fig:mhd1} and \ref{fig:mhd2}. A
filling factor of 50--75\% and a patch size of 1/8--1/6 of the PWN radius
($R_{\rm PWN}$) appear to give the best match between the simulations and the
data, such that they have comparable numbers of small-scale features and
contrast in the intensity maps. In Figure~\ref{fig:mhd3} we show the simulated
polarization maps. A large filling factor or small patch size generally give a
lower polarization fraction. This is because, as there are more patches with
different $B$-field orientation along a line of sight, depolarization becomes
more severe. We find that the same range of filling factor and patch size as
above provide the best fit to the observation. If this simple model is true,
then the simulation results suggest that mixing in this PWN is significant and
the turbulence scale is about $R_{\rm PWN}/8$ -- $R_{\rm PWN}/6$. The former
conclusion is in line with the speculation of \cite{temim15}, and should be
taken into account for further theoretical work with the Snail. Finally, we
note that the randomly oriented magnetic field in this simple model does not
satisfy the solenoid condition (i.e., $\nabla\cdot B = 0$). Therefore, it fails
to reproduce the magnetic loops in the PWN interior and the tangential field
structure at the boundary as observed. Instead, it is more useful to compare
the coherence scales of the models with the observations.

\subsection{Radio Spectrum of the Snail}
We determined a spectral index of $\alpha = -0.3 \pm 0.1$ for the Snail between
36 and 6\,cm. Such a flat spectrum is typical for PWNe. The flux density at
3\,cm is not well determined as the observations are only sensitive to smaller
structures in the PWN but not the whole nebula. The head of the Snail has a
steeper spectral index of $\alpha = -0.6 \pm 0.1$. This is unexpected, as X-ray
observations suggest that the particles are injected from the head
\citep[e.g.][]{sun99,temim09,temim15}, and therefore the materials in the head
should be younger than those of the body. This means that the spectral
difference probably cannot be explained by synchrotron cooling. We also note
that this spectral index is considerably steeper than the generally expected
value for PWNe ($-0.3 \leq \alpha \leq 0$), which is rather unusual and
requires further theoretical modeling to explain.

\subsection{Minimum-energy Magnetic Field Strength in the PWN}
\label{sec:bfield}
We computed the minimum-energy magnetic field strength by imposing the
conventional assumption, i.e.\ that the total energy $U_{\rm total}$ is
distributed between magnetic energy $U_B$ and particle energy $U_p$ in a way
that $U_{\rm total}$ is minimized. From synchrotron theory
\citep{pacholczyk70}, we have
\begin{equation} 
U_{\rm total} = U_B + U_p = \frac{B^2}{8\pi}\Phi V + c_{12}(1 + \eta) L_{\rm
syn} B^{-3/2}{\rm ,} 
\end{equation}
where $\Phi$ is the volume filling factor of the magnetic field, $V$ is the
volume of the emission region, $c_{12}$ is a constant weakly depending on the
spectral index and frequency range considered, $\eta$ is the energy ratio of
ions to electrons, and $L_{\rm syn}$ is the synchrotron luminosity of the PWN.
Minimizing $U_{\rm total}$ gives a minimum-energy field,
\begin{equation} 
B_{\rm m} = \left[ 6\pi c_{12} (1 + \eta) L_{\rm syn} \Phi^{-1} V^{-1}
\right]^{2/7}{\rm ,} \label{eqn:beq} 
\end{equation}
which is also referred to as the ``equipartition field'' in some literature. We
integrated the simple power-law radio spectrum for the entire PWN (as in the
fit of Figure~\ref{fig:radiospec}) between $10^7$ and $10^{13}\,{\rm Hz}$
\citep[e.g.][]{ng10,ng12}, and found $L_{\rm syn} = 2.9 \times 10^{35}
d_9^2\,{\rm erg\,s}^{-1}$, where $d_9$ is the distance to the Snail in units of
$9\,{\rm kpc}$. To estimate the volume $V$, we modeled the body as a sphere of
diameter $4^\prime$ and the head as a cylinder with a diameter of $0\farcm4$ and
a length of $1\farcm5$. The volume of the PWN is then $V = 1.8 \times 10^{58}
d_9^3\,{\rm cm}^3$. Substituting these parameters into Equation~\ref{eqn:beq}
gives
\begin{equation} 
B_{\rm m} = 36 (1 + \eta)^{2/7} \Phi^{-2/7} d_9^{-2/7}\, {\rm \mu G.}
\end{equation}
Assuming $\eta = 0$, $\Phi = 1$, and $d_9 = 1$, we find a minimum-energy field
of $36\,{\rm \mu G}$ in the Snail, which is of the same order of magnitude as
that estimated by modeling the broadband emission of the PWN
\citep[$11\,\mu{\rm G}$;][]{temim15}. Note that the computed minimum-energy
field strength can depend on the frequency range chosen. For instance, if we
only integrate up to $10^{11}\,{\rm Hz}$ instead of $10^{13}\,{\rm Hz}$, then
we obtain a slightly lower value of $26\,\mu{\rm G}$, and it is much less
sensitive to the lower frequency limit. It is also worth noting that the energy
in the pulsar wind is believed to be particle-dominated upon injection, and
hence the minimum-energy condition is not guaranteed to hold.

We also computed the minimum-energy field strength of the body and the head
separately, using identical procedures as above. For the body, we have $L_{\rm
syn, body} = 2.1 \times 10^{35}\,{\rm erg\,s}^{-1}$ and $V_{\rm body} = 1.8
\times 10^{58}\,{\rm cm}^3$, which gives $B_{\rm m, body} = 34\,\mu{\rm G}$.
For the head, we have $L_{\rm syn, head} = 4.1 \times 10^{33}\,{\rm
erg\,s}^{-1}$ and $V_{\rm head} = 9.9 \times 10^{55}\,{\rm cm}^3$, which gives
$B_{\rm m, head} = 69\,\mu{\rm G}$. The higher minimum-energy field strength
in the head as compared to that in the body could suggest that the reverse
shock has compressed the former but not yet the latter.

\subsection{Multiwavelength Comparison}
Figure~\ref{fig:g327}b shows that the radio peak of the head is offset from the
X-ray point source by $\sim 18^{\prime\prime}$ (corresponding to $0.8\,{\rm
pc}$ at a distance of $9\,{\rm kpc}$), which is larger than the astrometric
uncertainty of $\sim 1^{\prime\prime}$ of \textit{Chandra} and the radio beam
size (${\rm FWHM} \sim 15^{\prime\prime}$). A similar offset is found in other
systems, including G319.9$-$0.7 \citep{ng10} and the Lighthouse Nebula
\citep{pavan14} powered by PSR J1101$-$6101 \citep{halpern14}, in which the
radio and X-ray peaks are displaced by $4\,{\rm pc}$ and $0.7\,{\rm pc}$,
respectively. On the other hand, there is no detectable radio emission at the
location of the X-ray prongs and arc of the Snail. This is analogous to the
X-ray jet-like structures seen in the Guitar Nebula \citep{hui07} and the
Lighthouse Nebula \citep{pavan14}. This could be the result of the diffusion of
high-energy particles, but the density is too low for the radio emission to be
detected.  We also note that the prong structures are aligned with the edge of
the radio head.

The broadband synchrotron spectra of PWNe often steepen at high frequency due
to synchrotron cooling. For a single population of particles injected into the
system, the spectral break frequency $\nu_b$ and magnetic field strength
$B$ could be used to estimate the PWN age $\tau$ by \citep{gaensler06}
\begin{equation} 
\tau = \left(\frac{\nu_b}{10^{21}\,{\rm
Hz}}\right)^{-1/2}\left(\frac{B}{10^{-6}\,{\rm G}}\right)^{-3/2}\,{\rm kyr.}
\end{equation}
Using the $\nu_{\rm b} = 10^{13}\,{\rm Hz}$ obtained through extrapolation of
radio and X-ray spectra (Figure~\ref{fig:radiospec}b) and the minimum-energy
field strength obtained above ($B_{\rm m} = 36\,\mu{\rm G}$) gives $\tau =
46\,{\rm kyr}$ for the Snail. If we instead use the magnetic field strength
estimate from modeling \citep[$11\,\mu{\rm G}$;][]{temim15}, then we acquire an
even larger $\tau = 270\,{\rm kyr}$. These values differ significantly from the
age estimates of $11$--$29\,{\rm kyr}$ from previous studies
\citep{sun99,bocchino03,temim09,temim15}. We argue that the simple estimate
from the spectral break may not reflect the true age of the system. The change
in spectral index by a simple extrapolation of the radio and X-ray spectra is
$\Delta \alpha \approx 0.9$. This suggests that the actual spectral energy
distribution (SED) of the system can be much more complicated than expected, as
the result of a broken power-law injection spectrum and enhanced synchrotron
burnoff induced by the reverse shock interaction \citep[see][]{temim15}. This
can be verified by future observations between the radio and X-ray bands.

Figure~\ref{fig:radiospec}b shows the broadband SED of the Snail from radio to
$\gamma$-rays, with the $\gamma$-ray data from \textit{Fermi} and H.E.S.S.
\citep{acero12,acero13}. While the TeV $\gamma$-ray spectrum from H.E.S.S.
appears to align with the X-ray spectrum, this is probably just coincidence
\citep[see][]{temim15}.

\section{CONCLUSION} \label{sec:conclusion}
In this paper, we presented radio observations of the Snail PWN in SNR
G327.1$-$1.1 using ATCA and compared the results with the previous MOST study
\citep{whiteoak96}. The PWN shows the same morphology in all the five bands
between 36 and 3\,cm, consisting of a main circular body believed to be the
``relic'' PWN and a protruding head. The 6 and 3\,cm polarization maps reveal a
highly ordered magnetic field structure, in contrast with the turbulent
interior with a tangled magnetic field structure expected for a reverse shock
crushed PWN. This may suggest that the characteristic turbulence scale is
larger than the radio beam size. To explore this scenario, we developed a toy
model to simulate the emission from a turbulent PWN and found that a
characteristic turbulence scale of $\sim R_{\rm PWN}/8$ -- $R_{\rm PWN}/6$ with
a filling factor of $50$--$75\%$ can most closely match the observations.

We showed that the magnetic field at the head of the Snail aligns with the
nebular elongation. This serves as a good example of subsonic cometary PWN
systems with magnetic field parallel to the tail. In addition, we discovered
filamentary structures in the body which could be magnetic loops. It is,
however, unclear whether they are formed by kink instabilities as suggested for
3C~58 and G54.1$+$0.3, or by turbulence during the reverse shock interaction.
We also determined a spectral index of $\alpha = -0.3 \pm 0.1$ for the overall
PWN in the observed radio bands, which gives a minimum-energy magnetic field
strength of $B_{\rm m} \approx 36\,\mu{\rm G}$. The radio spectrum of the head
appears to be steeper than that of the body, which is difficult to explain
through synchrotron cooling.

\acknowledgements
The authors thank J.\ Lim for fruitful discussions. We also thank an anonymous
referee for helpful comments and suggestions which improved the paper. The
Australia Telescope Compact Array is part of the Australia Telescope National
Facility which is funded by the Commonwealth of Australia for operation as a
National Facility managed by CSIRO. MOST is operated by The University of
Sydney with support from the Australian Research Council and the Science
Foundation for Physics within the University of Sydney. This work is supported
by ECS grant of Hong Kong Government under HKU 709713P. P.O.S.\ acknowledges
partial support from NASA Contract NAS8-03060.

\textit{Facilities:} \facility{ATCA}, \facility{Molonglo Observatory}.

\end{document}